\documentclass[conference]{IEEEtran}
\usepackage{float}
\IEEEoverridecommandlockouts

\usepackage{cite}
\usepackage{amsmath,amssymb,amsfonts}
\usepackage{algorithm}
\usepackage{algpseudocode}
\usepackage{graphicx}
\usepackage{textcomp}
\usepackage{xcolor}
\usepackage{hyperref}
\usepackage{booktabs}
\usepackage{multirow}
\usepackage{tablefootnote} 
\usepackage{svg}
\usepackage{threeparttable}
\usepackage{bm} 
\usepackage{diagbox}
\def\BibTeX{{\rm B\kern-.05em{\sc i\kern-.025em b}\kern-.08em
    T\kern-.1667em\lower.7ex\hbox{E}\kern-.125emX}}
\begin{document}

\title{
Depth-Sequence Transformer (DST) for Segment-Specific ICA Calcification Mapping on Non-Contrast CT\\

}


\author{%
  \IEEEauthorblockN{%
    Xiangjian Hou\IEEEauthorrefmark{1}\IEEEauthorrefmark{2}\hspace{0.8em}%
    Ebru Yaman Akcicek\IEEEauthorrefmark{2}\hspace{0.8em}%
    Xin Wang\IEEEauthorrefmark{3}\\
    Kazem Hashemizadeh\IEEEauthorrefmark{2}\hspace{0.8em}%
    Scott Mcnally\IEEEauthorrefmark{2}\hspace{0.8em}%
    Chun Yuan\IEEEauthorrefmark{2}\hspace{0.8em}%
    Xiaodong Ma\IEEEauthorrefmark{2}%
  }\\[0.6ex]  

  \IEEEauthorblockA{\IEEEauthorrefmark{1}Dept.\ of Electrical \& Computer Engineering, University of Utah, Salt Lake City, UT, USA}%
  \IEEEauthorblockA{\IEEEauthorrefmark{2}Dept.\ of Radiology \& Imaging Sciences, University of Utah, Salt Lake City, UT, USA}%
  \IEEEauthorblockA{\IEEEauthorrefmark{3}Dept.\ of Electrical \& Computer Engineering, University of Washington, Seattle, WA, USA}\\[0.4ex]

  \texttt{\{xiangjian.hou,\,ebru.akcicek,\,s.hashemizadehkolowri\}@utah.edu}\\
  \texttt{\{scott.mcnally,\,chun.yuan,\,xiaodong.ma\}@hsc.utah.edu}\\
  \texttt{xwang99@uw.edu}%
}

\maketitle

\begin{abstract}

While total intracranial carotid artery calcification (ICAC) volume is an established stroke biomarker, growing evidence shows this aggregate metric ignores the critical influence of plaque location, since calcification in different segments carries distinct prognostic and procedural risks. However, a finer-grained, segment-specific quantification has remained technically infeasible. Conventional 3D models are forced to process downsampled volumes or isolated patches, sacrificing the global context required to resolve anatomical ambiguity and render reliable landmark localization.
To overcome this, we reformulate the 3D challenge as a \textbf{Parallel Probabilistic Landmark Localization} task along the 1D axial dimension. We propose the \textbf{Depth-Sequence Transformer (DST)}, a framework that processes full-resolution CT volumes as sequences of 2D slices, learning to predict $N=6$ independent probability distributions that pinpoint key anatomical landmarks.
Our DST framework demonstrates exceptional accuracy and robustness. Evaluated on a 100-patient clinical cohort with rigorous 5-fold cross-validation, it achieves a Mean Absolute Error (MAE) of \textbf{0.1 slices}, with \textbf{96\%} of predictions falling within a $\pm1$ slice tolerance. Furthermore, to validate its architectural power, the DST backbone establishes the best result on the public Clean-CC-CCII classification benchmark under an end-to-end evaluation protocol.
Our work delivers the first practical tool for automated segment-specific ICAC analysis. The proposed framework provides a foundation for further studies on the role of location-specific biomarkers in diagnosis, prognosis, and procedural planning. Our code available \href{https://github.com/Arctic-Xiangjian/DST-3D_model_for_medical_image_classification}{here}

\end{abstract}

\begin{IEEEkeywords}
Intracranial Arterial Calcification, Landmark Localization.
\end{IEEEkeywords}

\section{Introduction}

Intracranial internal carotid-artery calcification (ICAC) is a routine incidental finding on non-contrast Computed Tomography (NCCT)~\cite{kockelkoren2018intracranial,kockelkoren2017computed}. Prospective population studies and meta-analyses show that a larger total ICAC burden independently predicts incident ischemic stroke and is associated with accelerated cognitive decline, even after adjustment for traditional risk factors~\cite{kathuveetil2025arterial,bos2015atherosclerotic,arenillas2011intracranial}.
The clinical relevance of ICAC has fuelled the development of fully automated deep-learning pipelines that segment calcium and output a single volumetric biomarker~\cite{bortsova2017segmentation,bortsova2021automated}. Recent systems tested on thousands of NCCTs achieve expert-level accuracy for binary ICAC masks and volume quantification, demonstrating technical maturity for aggregate analysis~\cite{bortsova2021automated}.

\begin{figure}[!htbp]
  \centering
  \includegraphics[width=0.6\linewidth]{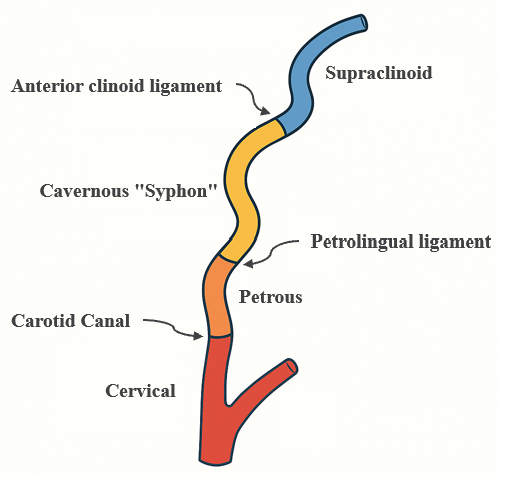}
\caption{Simplified four‐segment scheme of the internal carotid artery (ICA) used in this study. 
From proximal to distal, the \textbf{(i) Cervical}, \textbf{(ii) Petrous}, 
\textbf{(iii) Cavernous (“syphon”)}, and \textbf{(iv) Supraclinoid} segments are distinctly 
color‐coded along the vessel centerline. Lines inside indicate axial reference planes, 
manually delineated by board‐certified neuroradiologists, that define the boundaries between 
adjacent segments.}

  \label{fig:ica_segments}
\end{figure}

These successful pipelines have established total ICAC volume as a powerful prognostic marker. However, this aggregate approach, by its nature, treats all calcifications as equal, masking the distinct clinical implications of plaque at different anatomical locations. A growing body of evidence now suggests that a segment-specific analysis is crucial for advancing on two parallel fronts: \textbf{(i)} refining long-term risk prediction beyond what total volume can offer, and \textbf{(ii)} informing immediate, high-stakes procedural and surgical planning.

The following evidence underscores why this is clinically valuable.
\textbf{Cervical (extracranial):} Circumferential calcification exceeding $50\%$ limits balloon/stent expansion during carotid-artery stenting and leaves a higher burden of residual stenosis at completion, directly undermining revascularization efficacy~\cite{elsayed2022impact}. 
%
\textbf{Petrous:} When dense calcification co-exists with the tight curvature of the bony canal, it can accentuate the “ledge effect”, making aspiration or large-bore thrombectomy catheters more likely to stall at the bend and increasing device passes~\cite{ghali2019flow}.
\textbf{Cavernous \& supraclinoid:} 
%
%
Large-scale vessel-wall imaging shows that near $77\%$ of all intracranial ICAC localises to the cavernous and supraclinoid segments, and it has been found that higher scores correlate with stiffer parent arteries, impaired perforator reserve (ophthalmic, PCoA, AChA) and a greater burden of deep or branch-embolic infarcts—even when total ICAC volume is moderate~\cite{yang2021understanding}. Therefore, underscoring a segment-specific pathophysiology that simple whole-artery volume metrics cannot capture.

Taken together, these data argue that “\textbf{where}” calcium resides along the ICA may dictate stent deployability, and endovascular access as strongly as “\textbf{how much}” calcium is present. 
Although calcification volume shows no hemispheric prognostic difference~\cite{berghout2025morphometry}, segment-level mapping still hinges on the model’s ability to reliably distinguish left from right ICA, because each segment’s landmarks are inherently side-specific.
This clinical imperative motivates the shift from aggregate ICAC volume toward \textbf{segment-specific} quantification proposed in the present work. 

\begin{figure}[!htbp]
    \centering
    \includegraphics[width=1\columnwidth]{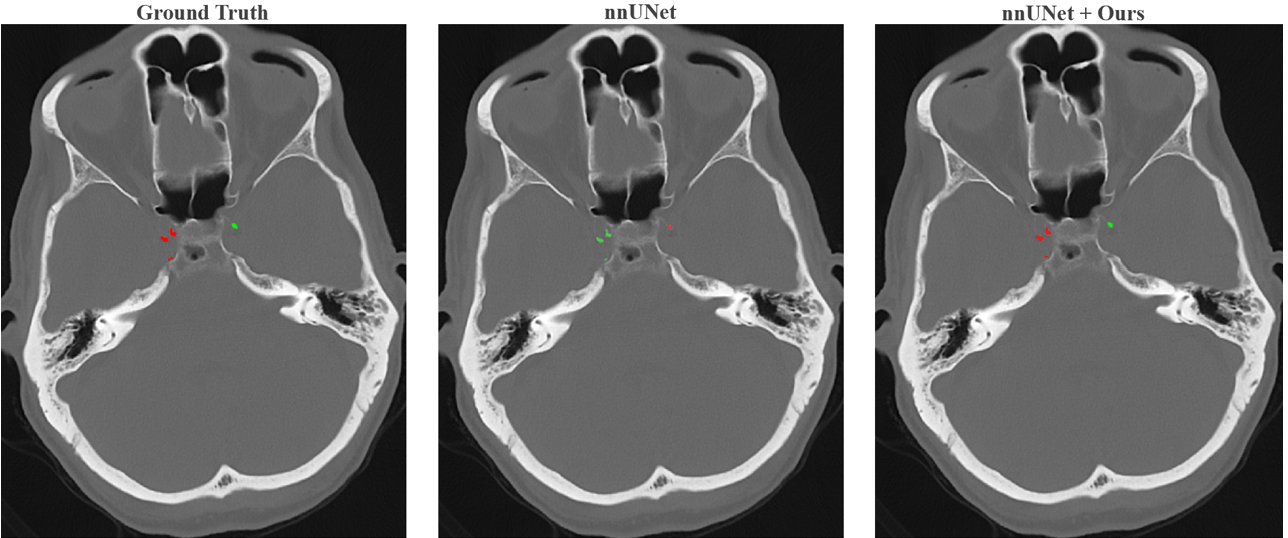} 
    \caption{
        \textbf{The Failure of Patch-Based Models in Tasks Requiring Global Context.}
        \textbf{(Left)} Ground truth labels for left (green) and right (red) ICA segments.
        \textbf{(Center)} A standard patch-based model (nnU-Net) catastrophically mislabels the hemispheres due to a lack of global context.
        \textbf{(Right)} Our proposed method correctly utilizes global information to assign anatomically correct labels.
    }
    \label{fig.compare}
\end{figure}

Yet, achieving this level of analysis has remained technically challenging. Manual segmentation is laborious for large-scale research. More fundamentally, a direct end-to-end approach is confounded by \textbf{perceptual aliasing}: the local texture and morphology of a calcified plaque are similar regardless of its anatomical segment. The identity of a plaque is therefore not defined by its appearance, but by its global anatomical context. This can lead to catastrophic localization errors; for instance, as vividly illustrated in Figure~\ref{fig.compare}, a standard patch-based model lacking this context frequently mislabels entire sections of the left carotid artery as the right, and vice-versa. This failure stems directly from the fundamental dilemma of conventional 3D models: to gain the necessary global context, they must drastically downsample the volume, which erases the landmark details they need to identify. Conversely, processing high-resolution patches preserves detail but sacrifices the long-range contextual cues required to resolve the perceptual ambiguity.

To overcome these limitations, we reformulate the 3D localization challenge. The goal is to identify the key \textbf{axial reference planes} that separates continuous anatomical segments. We simplify this complex 3D problem into a tractable one-dimensional task: \textbf{pinpointing the single, discrete slice index} that best represents each continuous boundary. To this end, we introduce the \textbf{Depth-Sequence Transformer (DST)}. 
%
%
Our work makes the following key contributions:
\begin{itemize}
    \item The first framework for automatic four-segment ICAC mapping on NCCT, validated on a 100-patient cohort with high accuracy and fidelity.
    \item An efficient architecture that achieves full-volume global context with a memory footprint two orders of magnitude smaller than standard 3D attention, from $O(H^2 W^2 D^{2})$ to $O(D^{2})$.
    \item A versatile architecture design, demonstrated by achieving state-of-the-art classification performance on the public Clean-CC-CCII benchmark.
\end{itemize}

\subsection{Related Work}

\textbf{ICAC Quantification and 3D Volumetric Analysis.} 
The clinical significance of ICAC’s spatial distribution, as opposed to just its total volume, is well-documented in manual radiological studies that have established a clear link between calcification in specific segments and distinct clinical outcomes~\cite{elsayed2022impact, yang2021understanding}. 
There are several deep-learning pipelines attempting ICA segment segmentation. However, they almost exclusively operate on high-contrast angiographic modalities such as CTA or time-of-flight MRA~\cite{nageler2023deep,chen2020automatedintracranialarterylabeling}—where the lumen is strongly opacified—rather than on routine NCCT in which vascular borders are barely discernible.
Automated analysis on NCCT has therefore been limited to quantifying the \textit{total} ICAC volume~\cite{bortsova2021automated}.
Therefore, a critical gap remains for establishing a tool that can automatically perform segment-specific mapping on routine NCCTs.

The dominant paradigm for 3D medical image analysis involves models designed for dense segmentation, such as 3D U-Nets and the Transformer-based successors like Swin UNETR~\cite{hatamizadeh2022unetr,hatamizadeh2021swin}. While highly effective for many tasks, their reliance on processing cropped patches makes them ill-suited for our landmark localization problem, as they lack the global context to resolve anatomical ambiguities. To address this, recent works specific to landmark detection, such as 
H3DE-Net~\cite{huang2025h3denetefficientaccurate3d}, have successfully employed sophisticated sparse attention to localize precise 3D coordinates. However, due to the difference in task formulation, direct comparison between H3DE-Net and our approach is not feasible.
%

In this work, we propose a different model called \textbf{DST}. Instead of designing a more complex attention module to operate within the 3D spatial domain, we reformulate the problem itself. Inspired by video transformers~\cite{gberta_2021_ICML,DBLP:journals/corr/abs-2103-15691} that linearize temporal frames, we linearize the depth axis into a slice sequence—an under-explored strategy in CT. Another way is to use pre-trained tokenizers~\cite{Jin_Diff3Dformer_MICCAI2024} as a feature extractor for downstream tasks. However, direct adaptations of these methods introduce their own challenges, such as high computational costs or the critical \textbf{feature alignment problem} when using frozen, general-purpose encoders~\cite{meng2024diffusionmodelactivationsevaluated}, and require a large amount of computational resources.

To reduce computational cost, we adapt the sequential processing paradigm but tailor it for \textit{static anatomical landmark localization}, featuring a lightweight, \textbf{end-to-end trained} architecture that efficiently processes variable-length sequences. This design leverages the simplicity of global attention on a 1D sequence while avoiding the feature alignment problems of pre-trained tokenizers, thereby providing a novel and effective solution.

\begin{figure*}[!htbp]
    \centering
    \includegraphics[width=0.95\linewidth]{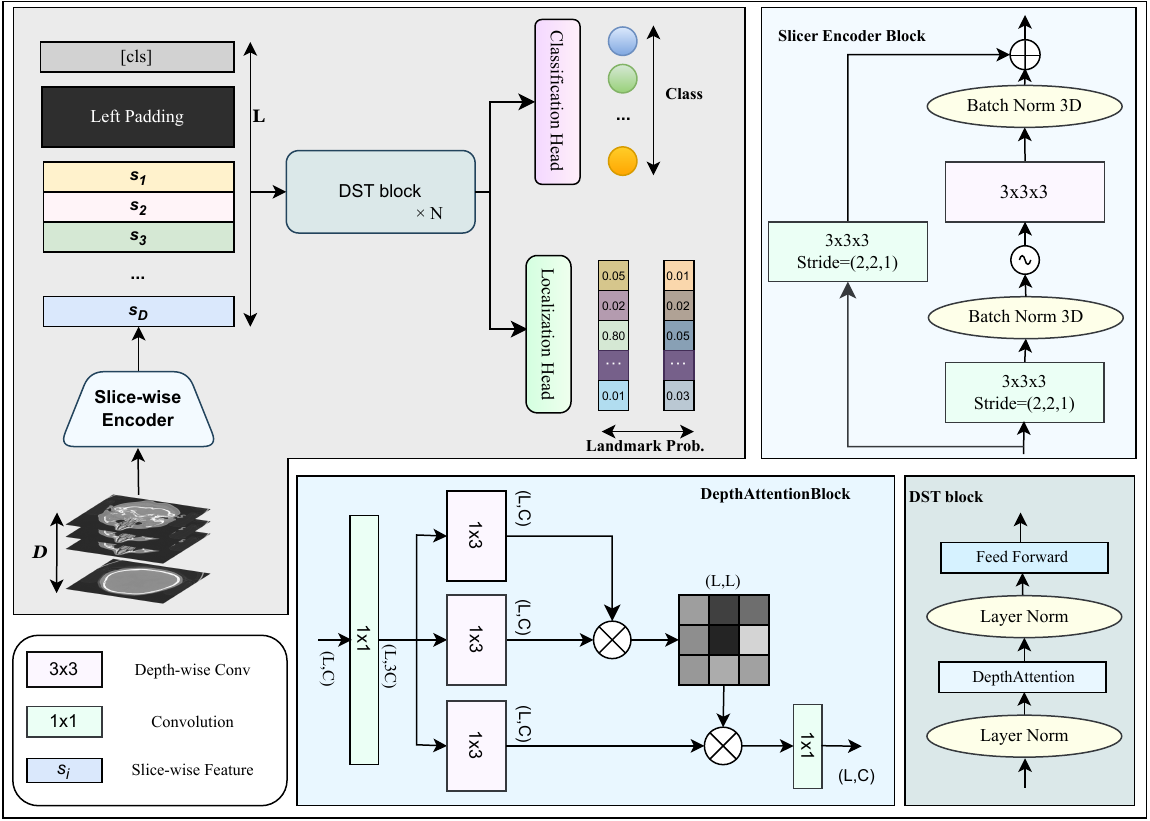}
    \caption{
    An overview of the proposed \textbf{Depth-Sequence Transformer (DST)} framework. 
    \textbf{(Left Panel)} The main pipeline: 
    (A)\textbf{Slice-wise Encoder} to generate a sequence of high-dimensional feature vectors, one for each axial slice.
    (B) `[cls]' token is added for classification tasks and \textbf{left-padded} to a fixed length.
    (C) A stack of N \textbf{DST blocks} processes the entire sequence using global self-attention.
    (D) Finally, two heads produce the outputs: a \textbf{Classification Head} operates on the `[cls]' token, while our primary \textbf{Localization Head} produces N independent probability distributions, each pinpointing one anatomical landmark.
    \textbf{(Right Panels)} Breakout diagrams of key modules: 
    The \textbf{Slice-wise Encoder Block} (top right) uses anisotropic 3D convolutions to preserve full axial resolution.
    The \textbf{DepthAttentionBlock} (bottom middle) illustrates our convolution-enhanced self-attention.
}
    \label{fig:model_architecture}
\end{figure*}

\section{Method}
Our work introduces the DST, a novel framework designed to perform automated segment-specific mapping of ICAC. Our approach builds upon the success of existing high-performance segmentation models, such as nnU-Net~\cite{isensee2021nnu}, which can accurately generate a binary mask of the total calcification burden.

The central challenge we address is how to assign a specific anatomical segment label to each calcified voxel within this pre-existing binary mask. To achieve this, our pipeline, illustrated in Figure~\ref{fig:model_architecture}, reformulates the 3D localization problem as a 1D sequence analysis task. It consists of two main stages:

First, a deterministic pre-processing algorithm is applied to the input 3D CT volume to establish an anatomical coordinate system and separate the left and right cerebral hemispheres, providing a crucial anatomical prior.

Second, our DST model takes the full-resolution CT volume as input to analyze the global anatomical context. It processes the volume as a sequence of axial slices and outputs the precise indices of the key anatomical landmarks that define the segment boundaries. These predicted indices are then used to assign the corresponding segment labels to all voxels within the initial binary mask. The following sections detail this approach.

\subsection{Anatomical Prior: Deterministic Hemisphere Separation}
A critical prerequisite for our task is the robust differentiation of the left and right carotid arteries. To this end, we developed a deterministic pre-processing pipeline to define the midsagittal plane for each CT volume, as detailed in Algorithm~\ref{alg:hemisphere_sep}.

Our approach combines anatomical morphology with DICOM metadata for a robust and reproducible result. We first isolate the skull using intensity thresholding and connected-component analysis to reliably determine the head's centroid. The orientation of the midsagittal plane is then defined directly using the direction cosines from the DICOM `ImageOrientationPatient' (0020,0037) tag. 


\begin{algorithm}[ht]
\small
\caption{Deterministic Hemisphere Separation}
\label{alg:hemisphere_sep}
\begin{algorithmic}[1]
    \State \textbf{Input:} 3-D CT volume $\mathcal{V}$, DICOM metadata $M$
    \State \textbf{Output:} Left mask $L_{mask}$, right mask $R_{mask}$
    \Statex \textit{1) Isolate Skull and Find Centroid}
    \State $S_{mask} \gets \text{FillHoles}(\text{LargestComponent}(\text{Threshold}(\mathcal{V})))$
    \State $\vec{c} \gets \text{Centroid}(S_{mask})$
    \Statex \textit{2) Define Midsagittal Plane from DICOM}
    \State $\vec{n} \gets \text{GetRowDirectionVector}(M)$ \Comment{From ImageOrientationPatient}
    \State $P_{plane} \gets \text{DefinePlane}(\text{point: } \vec{c}, \text{normal: } \vec{n})$
    \Statex \textit{3) Split Volume}
    \State $(L_{mask},\,R_{mask}) \gets \text{SplitVolumeByPlane}(\mathcal{V},\,P_{plane})$
    \State \textbf{return} $L_{mask},\,R_{mask}$
\end{algorithmic}
\end{algorithm}

\subsection{Problem Formulation as Parallel Probabilistic Landmark Localization}
We reformulate the 3D anatomical localization challenge as learning to predict $N$ independent, one-dimensional probability distributions, one for each of the $N=6$ key anatomical landmarks.

Let $\mathcal{V} \in \mathbb{R}^{H\times W\times D}$ be the input CT volume and $\{z_1, z_2, \dots, z_N\}$ be the ground truth slice indices of the $N$ landmarks. For each landmark $j \in \{1, \dots, N\}$, the ground truth $\mathcal{Y}^{(j)}$ is a one-hot vector of length $D$. Its $i$-th element is defined by the Kronecker delta, $\delta_{i,z_j}$, which is 1 if $i = z_j$ and 0 otherwise:
\begin{equation}
    \mathcal{Y}_i^{(j)} = \delta_{i, z_j} \qquad \forall i \in \{1, \dots, D\}
\end{equation}
Our model, a function $f_{\theta}$, maps the input volume $\mathcal{V}$ to a set of $N$ probability distributions, $\mathcal{P} = \{\mathcal{P}^{(1)}, \mathcal{P}^{(2)}, \dots, \mathcal{P}^{(N)}\}$, where each $\mathcal{P}^{(j)} \in \mathbb{R}^{D}$ represents the predicted probability distribution for the $j$-th landmark across all $D$ slices. This approach effectively decouples the difficult multi-point localization problem into $N$ simpler, parallel localization tasks.

This formulation advantage is that by localizing a few key indices across the entire volume, our model effectively sidesteps the challenges of extreme data sparsity and inter-class ambiguity that plague dense, multi-class segmentation approaches. This is made possible by our model's ability to process the entire volume at once, which is detailed next.

\subsection{Slice-wise Feature Encoder}
The initial stage of our model is a 3D CNN backbone designed to convert the input volume into a sequence of feature vectors.
A key design choice is the use of 3D convolutions with a stride of $(2, 2, 1)$, which aggressively downsamples the in-plane spatial dimensions while preserving the full resolution along the depth axis ($D$).
After several such blocks, a final spatial pooling layer collapses the spatial dimensions, producing a feature map of shape $(C, D)$. This can be interpreted as a sequence of $D$ feature vectors, where each $ C$-dimensional vector serves as a rich representation of its corresponding axial slice, aggregated from the full in-plane context.

\subsection{Sequence Preparation for Transformer}
To handle variable-length inputs from different CT scans, we chose a padding-based approach over resampling. While resampling is common, it distorts original data and, more critically, destroys the consistent physical relationship between slice indices and anatomical locations across patients, which is essential for a precise localization task. Our padding-and-masking strategy, in contrast, preserves full data fidelity and a consistent axial coordinate system. Specifically, we employ a \textbf{left-padding} strategy to bring all sequences to a fixed maximum length, a non-trivial choice shown to be crucial for maintaining correct positional alignment in sequence models~\cite{wang2023shallpretrainautoregressivelanguage}. A corresponding boolean attention mask is then created to ensure these padding tokens are ignored during the self-attention calculation, a practice demonstrated to be mathematically sound for Transformer-based architectures~\cite{merrill2025exact}.

\subsection{Dual Output Heads and the ``Attention Sink" Phenomenon}
The final output tokens from the Transformer are passed to two specialized heads:
\begin{itemize}
    \item \textbf{Parallel Landmark Localization Head:} For our primary task, a linear layer is applied to each of the $D$ output slice-tokens. This layer projects the feature dimension of each token into $N$ logits, one for each of the $N=6$ anatomical landmarks, resulting in a tensor of shape $(B, D, N)$. A Softmax function is then applied independently along the depth dimension ($D$) for each of the $N$ landmark channels. This yields the $N$ desired probability distributions, $\{\mathcal{P}^{(1)}, \dots, \mathcal{P}^{(N)}\}$, where each distribution $\mathcal{P}^{(j)}$ indicates the model's predicted location for the $j$-th landmark.
    \item \textbf{Classification Head:} For external validation, an optional classification head is applied to the output embedding of the prepended `[CLS]' token. Recent empirical studies have identified an ``attention sink" phenomenon in large language models, where attention scores naturally concentrate on initial tokens, such as the `[CLS]' token~\cite{gu2024attention,barbero2025llms}. This makes the `[CLS]' token a natural and effective aggregator of global information from the entire sequence, ideally suited for a global classification task.
\end{itemize}

\subsection{Training Objective}
Our framework's two output heads are trained independently for their respective tasks, using standard loss functions.

\subsubsection{Objective for Landmark Localization}
For our primary task of localizing the $N$ anatomical landmarks on the clinical ICAC dataset, only the Parallel Landmark Localization Head is active. The model is trained to minimize the sum of the individual Cross-Entropy losses for each of the $N$ predicted probability distributions. This composite localization loss, $\mathcal{L}_{loc}$, is defined as:
\begin{equation}
    \mathcal{L}_{loc} = \sum_{j=1}^{N} \mathcal{L}_{CE}\left(\mathcal{P}^{(j)}, z_j\right)
\end{equation}
where $\mathcal{L}_{CE}$ is the standard Cross-Entropy loss function, $\mathcal{P}^{(j)}$ is the predicted probability distribution over the slices for the $j$-th landmark, and $z_j$ is its ground truth slice index.

\subsubsection{Objective for External Validation}
When validating the architecture's general-purpose capability on the public Clean-CC-CCII~\cite{Heetalbenchmark} classification benchmark, only the Classification Head is utilized. The entire model is trained end-to-end on this separate task using a standard Cross-Entropy loss, $\mathcal{L}_{cls}$, applied to the logits produced by the `[CLS]' token.


\begin{figure*}[!htbp]
    \centering
    \includegraphics[width=0.95\linewidth]{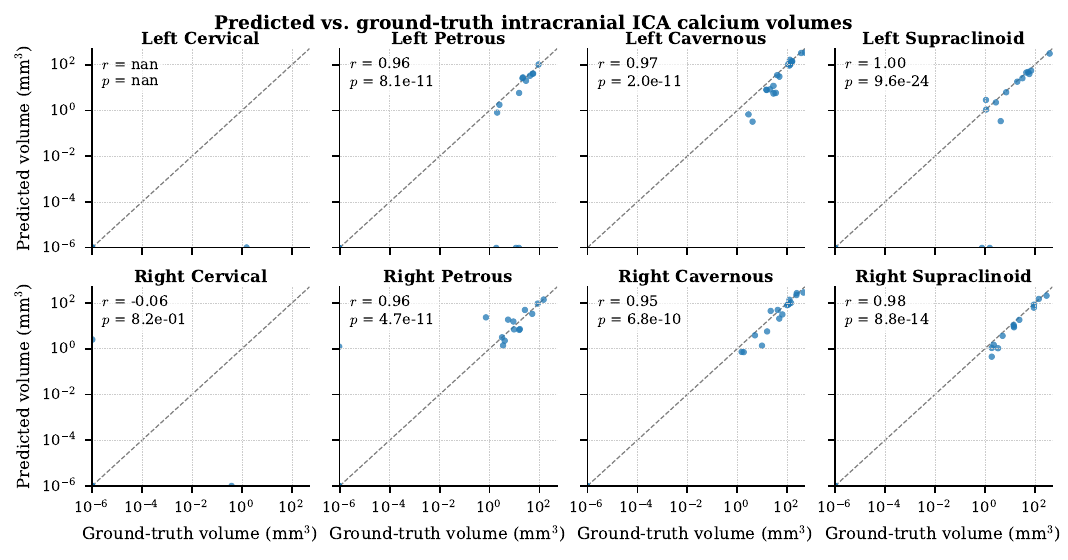} 
    \caption{
    \textbf{Excellent correlation between predicted and ground-truth calcium volumes for all intracranial segments.}
    Each scatter plot represents one of the eight ICA segments, plotting the predicted volume (y-axis) against the ground-truth volume (x-axis) on a log-log scale. The dashed line indicates perfect agreement.
    Our DST-based pipeline demonstrates very high and statistically significant Pearson correlations ($\textit{p} < 0.001$) for all six \textbf{intracranial} segments.
    Note that, no meaningful correlation is observed for the \textbf{extracranial} Cervical segments; as this anatomical region was at the edge of the scan's field-of-view, leading to near-zero calcification volumes for most subjects.
}
    \label{fig:gt_vs_pre}
\end{figure*}

\section{Experiments}

\subsection{Datasets}
We evaluated our proposed framework on two distinct datasets: a private clinical cohort for our primary task of ICA landmark localization, and a large public benchmark for validating the general-purpose capabilities of our DST architecture.

\textbf{In-house NCCT Dataset.} For our primary task, we retrospectively collected a dataset of 100 non-contrast head CT scans from our institution's patient database; 1 case was dropped because of the wrong gantry angle. The selection was performed randomly, with age range $40 \sim 90$, and average $72.5\pm10.8$. The ground truth for this study was established through a dual-reader process in which two trained experts independently annotated both the binary calcification masks and the six key landmark indices.


\textbf{Clean-CC-CCII~\cite{Heetalbenchmark}.} To validate the performance and generality of our DST architecture, we use the public China Consortium for Chest CT Image Investigation (CC-CCII) dataset, which contains over 3,393 CT images from 2,698 patients across three classes: novel coronavirus pneumonia (NCP), common pneumonia (CP), and normal controls (Normal). 


\subsection{Evaluation Metrics}
For our primary landmark localization task, we established a comprehensive suite of metrics. For each of the $N=6$ landmarks, let $z_{\text{true}}$ be the ground truth and $\mathcal{P}^{(j)}$ be the predicted probability distribution for landmark $j$. The final prediction $z_{\text{pred}}^{(j)}$ is the slice index with the highest probability.
\begin{itemize}
    \item \textbf{Mean Absolute Error (MAE):} Our primary metric for localization accuracy, measured in slices, defined as $\frac{1}{N}\sum_{j=1}^{N} |z_{\text{pred}}^{(j)} - z_{\text{true}}^{(j)}|$.
    \item \textbf{Top-k Accuracy:} The percentage of predictions where the true landmark slice is within the model's top $k$ most probable candidates. We report for $k=1$ and $k=2$.
    \item \textbf{Tolerance Accuracy ($\text{Acc}_{\tau=1}$):} The percentage of predictions with an absolute error of no more than 1 slice, reflecting clinical acceptability.
    \item \textbf{Quadratic Weighted Cohen's Kappa ($\kappa$):} Measures agreement by penalizing large errors more heavily than small errors.
\end{itemize}
For the Clean-CC-CCII classification task, we report standard metrics: overall Accuracy, macro-averaged F1-score, and AUC.


\subsection{Implementation Details and Evaluation Protocol}

For the definitive evaluation of our proposed DST model, both on our primary in-house landmark localization task and on the public Clean-CC-CCII classification benchmark, we employed a strict 5-fold patient-level cross-validation protocol. In contrast, the baseline segmentation experiments, intended to efficiently demonstrate the limitations of conventional end-to-end approaches, were conducted on a single, fixed data split ($70\%$ train, $10\%$ validation, $20\%$ test) for analysis.

In each fold, the patient cohort was partitioned to ensure that all scans from a single individual were confined to only one of the train $70\%$, validation $10\%$, or test $20\%$ sets. The validation set was used for model selection via early stopping. The final reported performance is the mean and standard deviation over the five distinct test folds. All models were trained using the SGD optimizer with a learning rate of 2e-3.

\subsection{Baselines for Comparison}
To demonstrate that a novel approach is necessary for our task, we first probe the limitations of the standard end-to-end segmentation paradigm. We selected the strongest possible representative of this paradigm as our baseline: the \textbf{3D full-resolution U-Net with a Residual Encoder}, implemented within the state-of-the-art segmentation framework~\cite{isensee2021nnu}. The continued dominance of this specific configuration as a leading general-purpose segmentation method has recently been rigorously validated across multiple datasets~\cite{isensee2024nnunetrevisitedrigorousvalidation}. This choice ensures we are comparing our pipeline against the most powerful and widely recognized conventional approach.

To efficiently establish its fundamental limitations on our fine-grained localization task, these baseline experiments were conducted on a single, fixed patient-stratified split.
\begin{enumerate}
    \item \textbf{Binary Segmentation:} A standard nnU-Net trained to produce a binary mask of all ICACs. This mask is then processed by our pipeline's outputs, and its high performance establishes a strong foundation.
    \item \textbf{End-to-End 2-Class Localization:} An nnU-Net trained to directly differentiate left vs. right ICA calcifications.
    \item \textbf{End-to-End 8-Class Segmentation:} An nnU-Net trained to directly segment all 8 distinct ICA classes.
\end{enumerate}
The inferior performance of baselines (2) and (3), detailed in the Results section, motivates our post-processing approach.

\section{Result}

\subsection{Performance on ICA Landmark Localization}

To demonstrate the limitations of conventional end-to-end methods and establish the necessity of our pipeline, we first conducted a series of baseline experiments on a fixed data split. The results are summarized in Table~\ref{tab:task_comparison}.

As expected, a 3D nnU-Net with Residual Encoder achieves decent performance on the simple binary segmentation task, yielding a Dice score of $0.66$. However, when tasked with a slightly more complex spatial task—differentiating between the left and right ICA—its performance degrades significantly. Furthermore, the model completely fails to converge when attempting the fine-grained 8-class segmentation task, confirming our hypothesis that patch-based models cannot resolve the anatomical ambiguity with sparse, visually similar classes.

In contrast, our pipeline, which builds upon the high-quality binary mask, shows clear superiority. Our deterministic algorithm robustly separates the hemispheres. 

\begin{table}[!htbp]
    \centering
    \caption{The failure of end-to-end approaches on fine-grained localization. Performance is measured by the Dice Score~($\uparrow$).}
    \label{tab:task_comparison}
    \resizebox{\columnwidth}{!}{
    \begin{tabular}{lccc}
        \toprule
        \diagbox{\textbf{Method}}{\textbf{Task}} & \textbf{Binary Seg.} & \textbf{Left vs. Right Seg.} & \textbf{8-Class Seg.} \\
        \midrule
        nnU-Net Res. Enc. (end-to-end) & 0.66 & 0.54 vs. 0.45 & Fails to converge \\[2pt]
        \midrule
        \multicolumn{4}{c}{\textit{Post-processing on nnU-Net's Binary Mask}} \\
        \midrule
        + Deterministic L/R Separation & 0.66 & 0.63 vs. 0.74 & - \\[2pt]
        + \textbf{DST (Ours)} & 0.66 &\textbf{0.63 vs. 0.74} & Fig.~\ref{fig:gt_vs_pre} \\
        \bottomrule
    \end{tabular}}
\end{table}

\begin{table*}[!htbp]
  \centering
  \caption{
        Detailed landmark localization performance of our proposed DST method.
    }
  \label{tab:metric_results}
  \begin{tabular}{lccc@{\hspace{6pt}}ccc}
    \toprule
    \multirow{2}{*}{\textbf{Metric}} &
    \multicolumn{3}{c@{\hspace{6pt}}}{\textbf{Left}} &
    \multicolumn{3}{c}{\textbf{Right}} \\
    \cmidrule(lr){2-4} \cmidrule(lr){5-7}
    & Carotid canal & Petrolingual ligament & Anterior clinoid ligament
    & Carotid canal & Petrolingual ligament & Anterior clinoid ligament \\
    \midrule
    MAE       & 0.06 $\pm$ 0.08 & 0.17  $\pm$ 0.12 & 0.15 $\pm$ 0.21 & 0.13 $\pm$ 0.17 & 0.17 $\pm$ 0.14 & 0.12 $\pm$ 0.12 \\[2pt]
    $\mathrm{Top}\text{-}1$ & 0.96 $\pm$ 0.05& 0.90 $\pm$ 0.06 & 0.93 $\pm$ 0.07 &0.94 $\pm$ 0.06  & 0.93 $\pm$ 0.05 &  0.93 $\pm$ 0.06\\[2pt]
    $\mathrm{Top}\text{-}2$ & 0.97 $\pm$ 0.04 & 0.90 $\pm$ 0.06 & 0.93 $\pm$ 0.07 & 0.94 $\pm$ 0.06 &  0.93 $\pm$ 0.05& 0.93 $\pm$ 0.06 \\[2pt]
    $\mathrm{Acc}_{\tau}$  & 0.99 $\pm$ 0.02 & 0.93 $\pm$ 0.06 & 0.96 $\pm$ 0.06 &   0.97 $\pm$ 0.04 &0.94 $\pm$ 0.06  & 0.97 $\pm$ 0.04 \\[2pt]
    $\kappa$          & 0.99 $\pm$ 0.01 & 0.99 $\pm$ 0.01 &  0.99 $\pm$ 0.01&0.98 $\pm$ 0.03  & 0.99 $\pm$ 0.01 & 1.00 $\pm$ 0.00 \\
    \bottomrule
  \end{tabular}
\end{table*}

The definitive performance of our proposed DST pipeline was evaluated using a rigorous 5-fold patient-level cross-validation. The detailed results for each of the six anatomical landmarks are presented in Table~\ref{tab:metric_results}.
Our DST module achieves an exceptionally low Mean Absolute Error (MAE) in localizing the 6 key landmarks, demonstrating the effectiveness of our proposed 1D sequence reformulation.

%
The overall Mean Absolute Error across all landmarks is 0.13 slices, with very low standard deviation, indicating high precision and stability. Critically for clinical application, we achieved almost $96\%$ accuracy ($\mathrm{Acc}_\tau =0.96$), where a prediction is considered correct if its error is no more than one slice. The near-perfect Top-1 and Top-2 accuracies, along with a Quadratic Weighted Kappa approaching 1.0, further underscore the model's reliability. 
Furthermore, the excellent agreement between predicted and ground-truth calcium volumes for all intracranial segments shown in Fig.~\ref{fig:gt_vs_pre} validates our framework as a highly accurate tool for generating meaningful, segment-specific quantitative analysis.

\begin{table}[!htbp]
  \centering
  \caption{Performance comparison on the Clean-CC-CCII benchmark against standard baseline methods.}
  \label{tab:model_comparison}

  \begin{threeparttable}
    \resizebox{\columnwidth}{!}{%
      \begin{tabular}{lcccc}
        \toprule
        \textbf{Model} & \textbf{Flops} (GFlops) & \textbf{AUC} & \textbf{F1} & \textbf{Acc} \\
        \midrule
        DenseNet121\_3D~\cite{Heetalbenchmark}
          & 21.45 & $0.98\pm0.01$ & $0.89\pm0.03$ & $0.90\pm0.02$ \\[2pt]
        ResNet101\_3D~\cite{Heetalbenchmark}
          & 60.91 & $0.99\pm0.00$ & $0.92\pm0.01$ & $0.92\pm0.01$ \\[2pt]
        ResNet50\_3D~\cite{Heetalbenchmark}
          & 40.89 & $0.99\pm0.00$ & $0.92\pm0.02$ & $0.92\pm0.02$ \\[2pt]
        EfficientNetv2\_3D~\cite{pmlr-v139-tan21a}
          & 33.63 & $0.98\pm0.03$ & $0.89\pm0.01$ & $0.89\pm0.01$ \\[2pt]
        Convnextv2\_3D~\cite{woo2023convnext}
          & 29.98 & $0.94\pm0.01$ & $0.83\pm0.03$ & $0.83\pm0.03$ \\[2pt]
        Video Swin Transformer~\cite{Liu_2022_CVPR}
          & 60.33 & $0.96\pm0.01$ & $0.85\pm0.01$ & $0.85\pm0.01$ \\[2pt]
        Diff3Dformer\tnote{1}~\cite{Jin_Diff3Dformer_MICCAI2024}
          & --    & $0.91$        & $0.84$        & $0.85$        \\[2pt]
        \midrule
        \textbf{DST (Ours)}
          & 35.57 & \boldmath$0.99\pm0.01$ & \boldmath$0.92\pm0.01$ & \boldmath$0.92\pm0.01$ \\[2pt]
        \bottomrule
      \end{tabular}
    }

    \begin{tablenotes}
      \footnotesize
      \item[1] Performance taken from the original paper.
    \end{tablenotes}
  \end{threeparttable}
\end{table}

\subsection{Validation of Architectural Generality on a Public Benchmark}



To validate the general-purpose power and scalability of our DST architecture, we benchmarked it against a comprehensive suite of standard models on the public Clean-CC-CCII dataset. As presented in Table~\ref{tab:model_comparison}, our DST framework demonstrates highly competitive performance with  greater computational efficiency. It requires only \textbf{35.57 GFLOPs}, a significant reduction compared to ResNet101 (60.91 GFLOPs) and more efficient than ResNet50 (40.89 GFLOPs). This favorable computational profile is further reflected in its requirement of just \textbf{800~MB} of CUDA memory and \textbf{10~ms} of inference time per 128×128×64 volume on an NVIDIA RTX A100 GPU.

\subsection{Ablation Studies}

To validate our design choices and understand the contribution of each key component within our DST framework, we conducted a series of comprehensive ablation studies. First, we evaluated the importance of the core components on our primary landmark localization task. Second, we assessed the impact of our architecture's depth on the general-purpose classification task to justify our model's capacity.

\subsubsection{Contribution of Core Components on Landmark Localization}

We began by systematically degrading or altering core elements of our model. The specific ablations were: \textbf{w/o Attention:} The stack of DepthAttentionBlocks was replaced with a sequence of standard 1D convolutional layers to verify the necessity of the self-attention mechanism. \textbf{Right-Padding:} The padding strategy was switched from left- to right-padding to empirically justify our design choice.

The results of our ablation study, presented in Table~\ref{tab:ablation_localization}, reveal the critical and distinct contribution of each of our key design choices. 
Most notably, replacing the self-attention mechanism with standard convolutions (\textit{w/o Attention}) leads to a significant degradation in localization performance, while the model can still provide a coarse localization signal (MAE $\approx$ 0.3 slices) owing to our strong CNN encoder.
%
Finally, the inferior results of the \textit{Right-Padding} experiment empirically validate our choice of left-padding. This strategy provides a crucial advantage by creating a \textbf{fixed anatomical anchor} at the end of the sequence, giving the Transformer a stable coordinate system to learn relative positions.



\begin{table}[!htbp]
    \centering
    \caption{Ablation study of core components on the ICA landmark localization task.}
    \label{tab:ablation_localization}
    \resizebox{\columnwidth}{!}{
    \begin{threeparttable}
    \begin{tabular}{l|cc|c} 
        \toprule
        \textbf{Metric} & \textbf{\textit{w/o} Attention} &  \textbf{Right-Pad} & \textbf{Full Model (Ours)} \\
        \midrule
        MAE (slices) $\downarrow$   & 0.2722  & 0.1457 & \textbf{0.1325} \\
        Top-1 Acc (\%) $\uparrow$     & 91.85   & 92.18  & \textbf{93.12}  \\
        Acc$_{\tau=1}$ (\%) $\uparrow$ & 93.97   & 95.66  & \textbf{95.79} \\
        \bottomrule
    \end{tabular}
    \end{threeparttable}}
\end{table}

\subsubsection{Impact of Architectural Depth on Classification}

To determine the optimal depth for our architecture, we evaluated its classification performance on the Clean-CC-CCII benchmark while varying the number of stacked DST blocks. As shown in Table~\ref{tab:ablation_depth}, when the Transformer blocks are removed entirely (0 layers), the model's classification performance collapses to near-random levels. This confirms that for global tasks like classification, there must be a mechanism to aggregate information from the entire sequence to the `[CLS]' token.


\begin{table}[!htbp]
    \centering
    \caption{Ablation study on the number of DST layers, evaluated on the 3-class Clean-CC-CCII task.}
    \label{tab:ablation_depth}
    \begin{tabular}{lccc}
        \toprule
        \textbf{Metric} & \textbf{0 Layers} & \textbf{1 Layer} & \textbf{8 Layers} \\
        \midrule
        AUC ($\uparrow$)      & 0.521 & 0.980 & \textbf{0.985} \\
        F1-score ($\uparrow$) & 0.338 & 0.899 & \textbf{0.917} \\
        Accuracy ($\uparrow$) & 0.345 & 0.904 & \textbf{0.919} \\
        \bottomrule
    \end{tabular}
\end{table}

\section{Discussion}

We proposed the \emph{Depth-Sequence Transformer} (DST) to tackle the long-standing problem of \textbf{segment-specific} ICA calcification mapping on NCCT. Our approach recasts the 3-D spatial localisation task as a 1-D sequence-labelling problem, effectively modeling long-range anatomical dependencies that are often missed by 3-D CNNs with limited receptive fields. This is achieved via DST's two-stage architecture: a lightweight CNN first extracts per-slice feature tokens, which are then processed by a Transformer to capture the full sequential context. This design allows our framework to (i) preserve full-resolution anatomical context, (ii) reduce the attention cost to $O(D^{2})$, and (iii) by providing precise anatomical landmarks, our framework complements strong segmentation baselines like nnU-Net, which can then excel at detailed segmentation without being confounded by large-scale localization ambiguities.



\subsection{Generalisability beyond ICA}

When trained on the Clean-CC-CCII dataset without architecture changes or task-specific optimizations, DST surpasses published baselines, underscoring its strength as a generic 3-D backbone, which means that the architecture is not over-specialized to vascular anatomy. We attribute this versatility to two factors: 

(1) the 1-D formulation keeps GPU memory constant w.r.t.\ in-plane resolution, enabling \emph{full-field} inference instead of down-sampling;  

(2) the hybrid CNN–Transformer split cleanly separates low-level feature extraction from high-level reasoning, a design pattern that could benefit vertebrae labeling, cardiac phase detection, or vessel bifurcation mapping.



\subsection{Limitations and future work}

\textbf{Single-centre data.} Although five-fold cross-validation mitigates overfitting, external validation on multi-centre cohorts is required.

\textbf{Outcome prediction.} Current experiments stop at landmark localisation. Coupling DST with longitudinal clinical endpoints (e.g.\ stroke, cognitive decline) could yield fully automated risk scores.

\textbf{Broader Applications.} Its success on disparate tasks motivates further validation on new public benchmarks and our planned contribution of DST to open-source model libraries.


In summary, DST's success across diverse tasks validates sequence analysis as a powerful and efficient paradigm for fine-grained volumetric understanding in medical imaging.

\bibliographystyle{IEEEtran}
\bibliography{IEEEabrv,IEEEexample,IEEEfull}

\end{document}